\documentclass[twocolumn,pra]{revtex4}
\usepackage[utf8]{inputenc}
\usepackage{color}
\usepackage{braket}
\usepackage[version=3]{mhchem}
\usepackage{amsmath}
\usepackage{graphicx} %package to manage images
\graphicspath{{./images/}}

\begin{document}
\title{Bath mediated decay of density waves in a disordered Bose lattice gas}
\author{Sridhar Prabhu}
\author{Erich J. Mueller}
\affiliation{Laboratory of Atomic and Solid State Physics, Cornell University, Ithaca, New York}
\date{June 2020}

\begin{abstract}
{Motivated by a recent experiment, we study the dynamics of bosons in a disordered optical lattice, interacting with a variably sized bath of disorder free atoms. As the number  of  particles in the bath is increased, there is a transition between ``localized'' and ``ergodic'' behavior, which are characterized  by the long-time behavior of an initial density wave.  We model the dynamics with a stochastic mean field theory, reproducing  the central observations of the experiment. A key conclusion from our study is that particle loss plays an important role.}
\end{abstract}

\maketitle
\section{Introduction}
Many-body localization (MBL), where interactions fail to thermalize an isolated and disordered quantum system, provides an opportunity to confront fundamental questions about quantum statistical mechanics \cite{Review1,Review2,Review3}. For example, localized systems violate the Eigenstate Thermalization Hypothesis \cite{ETH1,ETH2,ETH3}. This lack of thermalization has been used as a signature of MBL \cite{MBL1,Main,Main-Disorder,MBL3}, and has been proposed as a tool for protecting quantum information \cite{QuantumInfo1,QuantumInfo2,QuantumInfo3}. 

As MBL requires a closed quantum system, a natural question is its stability in the presence of an external bath \cite{Bath1,Bath2,Bath3,Bath4}. In this paper we model the dynamics of a disordered interacting system with a varying bath size. We follow the experimental protocol in Ref \cite{Main}, which uses a two-component Bose gas, where one of the two components is sensitive to disorder. The disorder insensitive component acts as a thermal bath. The experiment finds a threshold bath size which separates localized and ergodic behavior. Our analysis captures this result. Surprisingly, we find that atom loss is important for this observation.

We use a stochastic variant of the Gutzwiller mean field theory \cite{Gutzwiller1, Gutzwiller2, Gutzwiller3, Gutzwiller4}. The Gutzwiller mean field theory can be viewed as a variational method based upon a product wavefunction ansatz. It has been very successful at calculating both the static and dynamic properties of bosons in optical lattices, including slow transport in the presence of disorder \cite{BGTheory1,BGTheory2}. Though a mean field theory is unable to capture some of the features of MBL, such as the growth of entanglement entropy or the behavior of individual energy eigenstates, it has been used to model similar experiments   to the one we are considering here \cite{GMFT-Main, Main-Disorder}. Ultimately, the agreement between our results and the experiment \cite{Main} shows that we have included the relevant physics in our modeling.

Due to background gas collisions, and 
%loss from 
inelastic light scattering,
the number of atoms falls with time.
To model these random atom loss events we use a stochastic wavefunction method
%The stochastic feature of our method allows us to account for random atom loss events 
\cite{Stochastic-Loss}.  
%The loss events come from
%This  allows us to  
%background gas collisions and 
%loss from 
%inelastic light scattering. 
We find that these  losses have a  significant impact on the lifetime of the density waves.  Importantly, we only see a threshold behavior when we include atom loss.

Our paper is organized as follows: in Section II, we introduce the two-component Bose Hubbard model which describes the experiment. In Section III, we outline our theoretical approach. Section IV details the results of our calculations. Section V summarizes our results.
%, and provides outlook on how our predictions can be verified by experiment.

\section{Model}\label{model}

In the experiment,  bosonic atoms hop on a 2D square lattice. The atoms can be in one of two internal states: $\ket{c}$ -- for clean, which experience a defect free lattice potential, and $\ket{d}$ -- for dirty, which, in addition to the lattice potential, also experience a random on-site potential. Experimentally \cite{Main}, these two components are different hyperfine states of the boson \ce{^{87}Rb}: $\ket{c} = \ket{F=1,m_{F}=-1}$, $\ket{d} = \ket{F=2,m_{F}=-2}$, and the disorder is applied optically, using a transition which affects only the $\ket{d}$ states.

Following \cite{Main}, we model this experiment with a Bose Hubbard Hamiltonian
\begin{equation} \label{BH Ham}
H = -J\!\!\! \sum_{\langle i,j \rangle,\sigma}\!\!\! \hat{a}_{i,\sigma}^\dagger \hat{a}_{j,\sigma}  +  \frac{U}{2} \sum_{i} \hat{n}_{i} (\hat{n}_{i} - 1)    +  \sum_{i}\delta_{i}\hat{n}_{i,d}
\end{equation}
Here $\hat n_i=\hat{n}_{i,c}+\hat{n}_{i,d}$ counts the number of particles on site $i$ ($\hat{n}_{i,c}=\hat{a}_{i,c}^\dagger \hat{a}_{i,c}$ is the number of clean particles on site $i$ and $\hat{n}_{i,d}=\hat{a}_{i,d}^\dagger \hat{a}_{i,d}$ is the number of dirty particles on site $i$).  In the sums, $\sigma \in \{c,d\}$, $i$ runs over all the sites of the system, and $\langle i,j\rangle$ represents nearest neighbor sites. The first term in Eq.~(\ref{BH Ham}) corresponds to %nearest-neighbor 
hopping with 
%an experimental 
tunneling rate $J = 2\pi$ x $24.8\hbar$ Hz.  The natural timescale of the system is $\tau = \hbar/J$. The intra- and inter-species repulsive interactions are roughly equal in strength and are characterized by a single parameter $U = 24.4J$. The disorder distribution $\delta_i$ is characterized by its full width at half maximum $\Delta=28J$ and a correlation length $\xi= 0.6a_{lat}$, where $a_{lat} = 532$nm is the lattice spacing.

\begin{figure}[tbp]
    \centering
    \includegraphics[width=\columnwidth]{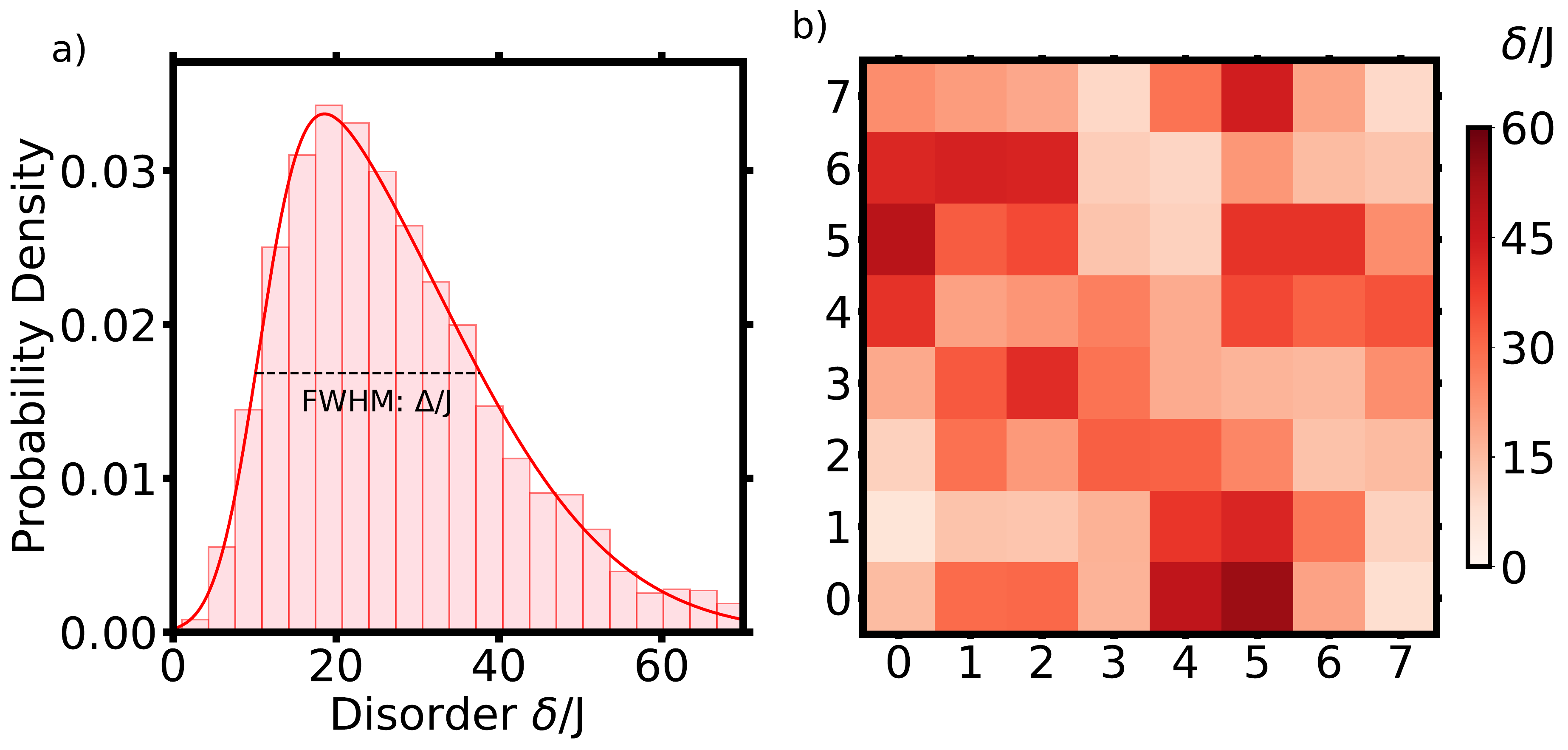}
    \caption{(Color Online) a) Probability density of the on-site disorder $\delta$, generated by the procedure in the text, normalized by the hopping matrix element $J$.  Also shown is the best fit skew-normal distribution. The strength of the disorder is measured by its full width at half maximum, $\Delta/J$. b) Example disorder realization used in our numerics, with 64 sites, separated by lattice-spacing $a$.  The correlation length is $\xi=0.6a$.}
        \label{Fig:Disorder}
\end{figure}

To construct disorder realizations, we use the protocol described in the appendix of \cite{Main}. On each site we sample an auxiliary variable $x$ from a uniform distribution $\in [0,W]$ and then square it. It is then convolved with a two dimensional gaussian, with periodic boundary conditions, to reproduce the observed correlation length. We choose $W$ 
so that the width of the resulting distribution matches the experimentally measured $\Delta$.  The resulting probability distribution is shown in Fig.~\ref{Fig:Disorder}, and is well approximated by a skew-normal distribution.

The system is prepared with a high contrast density wave where alternate columns are either occupied or empty (see call-outs of Fig.~\ref{Fig:F=1}). The initially occupied sites are referred to as ``even" and the empty ones as ``odd". Using a coherent transition, the experiment puts each atom into a superposition of the two hyperfine states. The number of particles on each sublattice,
$N_{\rm even}$ and $N_{\rm odd}$, are monitored as a function of time, and are used to construct the imbalance, 
\begin{equation}
    I= \frac{N_{\rm even}-N_{\rm odd}}{N_{\rm even}+N_{\rm odd}}.
\end{equation}
The experiment has an initial imbalance  of $I=0.91 \pm 0.01$ with $N=124 \pm 12$ particles. The imbalance decreases as a function of time, and is a proxy for equilibration.  

Also important to our analysis,
the system experiences particle loss with a characteristic time of $\Gamma^{-1} = (3300 \pm 300) \tau$.  This loss rate appears to be the same for both components. Importantly, we are only able to reproduce the experimental observations when we include this loss in our modelling. 

The experiment is repeated several times, with the quoted imbalance averaged over multiple realizations of the disorder. The experiment is then repeated for various values of clean particles, $N_c$, with the total number of particles $N=N_c+N_d$ fixed. In our numerics we follow this same procedure, and take $N$ to be exactly half the number of sites.

\section{Method}

Our modelling is based upon the Gutzwiller mean field approach \cite{Gutzwiller1}, which will be outlined in Section~\ref{mft}. To include loss, we use a stochastic  wavefunction approach, described in Section~\ref{sw}.  This combination of the Gutzwiller meanfield theory with the stochastic wavefunction approach is novel.

\subsection{Mean Field Theory}\label{mft}

The Gutzwiller mean field theory
can be viewed as a variational method in which one makes the product state ansatz, $\ket{\psi} = \prod_{i} \ket{\psi_{i}}$, where $i$ indexes the lattice sites. The local wavefunction can be written in the occupation basis as:
\begin{equation}
    \ket{\psi_{i}} = \sum_{n_{c}, n_{d} = 0}^{\infty} f_{n_{c},n_{d}}^{i} \ket{n_{c},n_{d}}     
\end{equation}
We truncate to $n_c + n_d = 3$.  We find negligible difference when we relax this cutoff.

The procedure described in Section~\ref{model} naturally yields an initial wavefunction of this product form with:
\begin{eqnarray}
\nonumber
\ket{\psi_e} &=& \sin\phi \ket{0,0} + \cos\phi\sin\theta \ket{1,0}+\cos\phi\cos\theta\ket{0,1}
\\
\nonumber
\ket{\psi_o} &=& \cos\phi \ket{0,0} +
\sin\phi\sin\theta \ket{1,0}+\sin\phi\cos\theta\ket{0,1}.
\end{eqnarray}
Here $\ket{\psi_e}$ and $\ket{\psi_o}$ correspond to $\ket{\psi_{i}}$ with $i$ representing a site on an even or odd column. The probability of having a particle on a given even or odd site is $\cos^2 \phi$ and $\sin^2\phi$ respectively. The fraction of clean particles is set by $\sin^2\theta$. We take the imbalance to be given by the experimental value $I_0=\cos(2\phi)=0.91$, and vary the clean fraction.

Following the procedure in \cite{GMFT_Procedure}, we construct equations of motion for the parameters by extremizing $S=\langle \psi |i \partial_t -H |\psi\rangle$. This yields
\begin{widetext}
\begin{multline}\label{master}
    i \frac{\partial f_{n_{c},n_{d}}^{i}}{\partial t} = -J\left(
    \sqrt{n_{c}}\, \phi_{i,c}^{*} f_{n_{c} - 1, n_{d}}^{i}  + \sqrt{n_{c} + 1} \,\phi_{i,c} f_{n_{c} + 1, n_{d}}^{i*} + \sqrt{n_{d}}\,\phi_{i,d}^{*} f_{n_{c}, n_{d} - 1}^{i}  + \sqrt{n_{d} + 1}\,\phi_{i,d} f_{n_{c}, n_{d} + 1}^{i*} \right) \\
    + \left(\frac{U}{2}(n_{c} + n_{d})(n_{c} + n_{d} - 1) + \delta_{i}n_{d}\right) f_{n_{c},n_{d}}^{i},
\end{multline}
\end{widetext}
where the mean-fields involve sums over the nearest neighbor sites,
\begin{eqnarray}
\phi_{i,c} &=& \sum_{\langle i,j \rangle} \sum_{n_{c},n_{d}} f_{n_{c},n_{d}}^{j*} f_{n_{c} - 1,n_{d}}^{j} \sqrt{n_{c}} \nonumber \\
\phi_{i,d} &=& \sum_{\langle i,j \rangle} \sum_{n_{c},n_{d}} f_{n_{c},n_{d}}^{j*} f_{n_{c},n_{d} - 1}^{j} \sqrt{n_{d}}.
\end{eqnarray}

\subsection{Stochastic Wavefunction Approach to Modelling Loss}\label{sw}

In the experiment, the atom number decays with a rate $\Gamma \sim 1/3300\tau$. This loss rate is %observed to be
the same for both the clean and dirty components. Sources of such atom loss include off-resonant light scattering from the optical lattice \cite{ORES1, ORES2}, and background gas collisions. We include these processes via the stochastic-wavefunction approach from Ref. \cite{Stochastic-Loss}. % We implement a split-step approach to capture the coherent and incoherent contributions.

We use jump operators proportional to $\hat a_{i,\sigma}$, resulting in a non-Hermitian contribution to the Hamiltonian,
\begin{equation}\label{loss}
    H_{Loss} =  - \frac{i\Gamma}{2} \sum_{i} (\hat{a}^{\dagger}_{c,i}\hat{a}_{c,i} + \hat{a}^{\dagger}_{d,i}\hat{a}_{d,i}).
\end{equation}
In a timestep $\delta t$, we first evolve the coefficients of the wavefunction via Eq.~(\ref{master}) using a single step of the Runge-Kutta (RK4) method. We use a projection technique, detailed in Appendix~\ref{lagrange}, to ensure that numerical errors do not lead to any artificial atom loss. Subsequently, we evolve the wavefunction via Eq.~(\ref{loss}) using a standard Euler step. This non-Hermitian evolution reduces the wavefunction norm,
\begin{equation}\label{particle_loss}
    \Braket{\psi(t+\delta t)|\psi(t+\delta t)} = 1 - \sum_i \left( \delta p_{c,i}+\delta p_{d,i}\right),
\end{equation}
where $\delta p_{c,i}$ and $\delta p_{d,i}$ are the probability of loosing a clean or dirty particle from site $i$. We then randomly determine if a loss event occurs. If a loss occurs at site $i$, we apply the corresponding annihilation operator: $|\psi(t+\delta t)\rangle \to  \hat a_{i,\sigma}|\psi(t+\delta t)\rangle$.  We then explicitly normalize the wavefunction and repeat for the next time step. Our implementation is further described in  Appendix~\ref{atom_loss}.

\section{Results}

\subsection{Imbalance Dynamics}\label{sec:id}

\begin{figure}[tbph]
    \centering
    \includegraphics[width=\columnwidth]{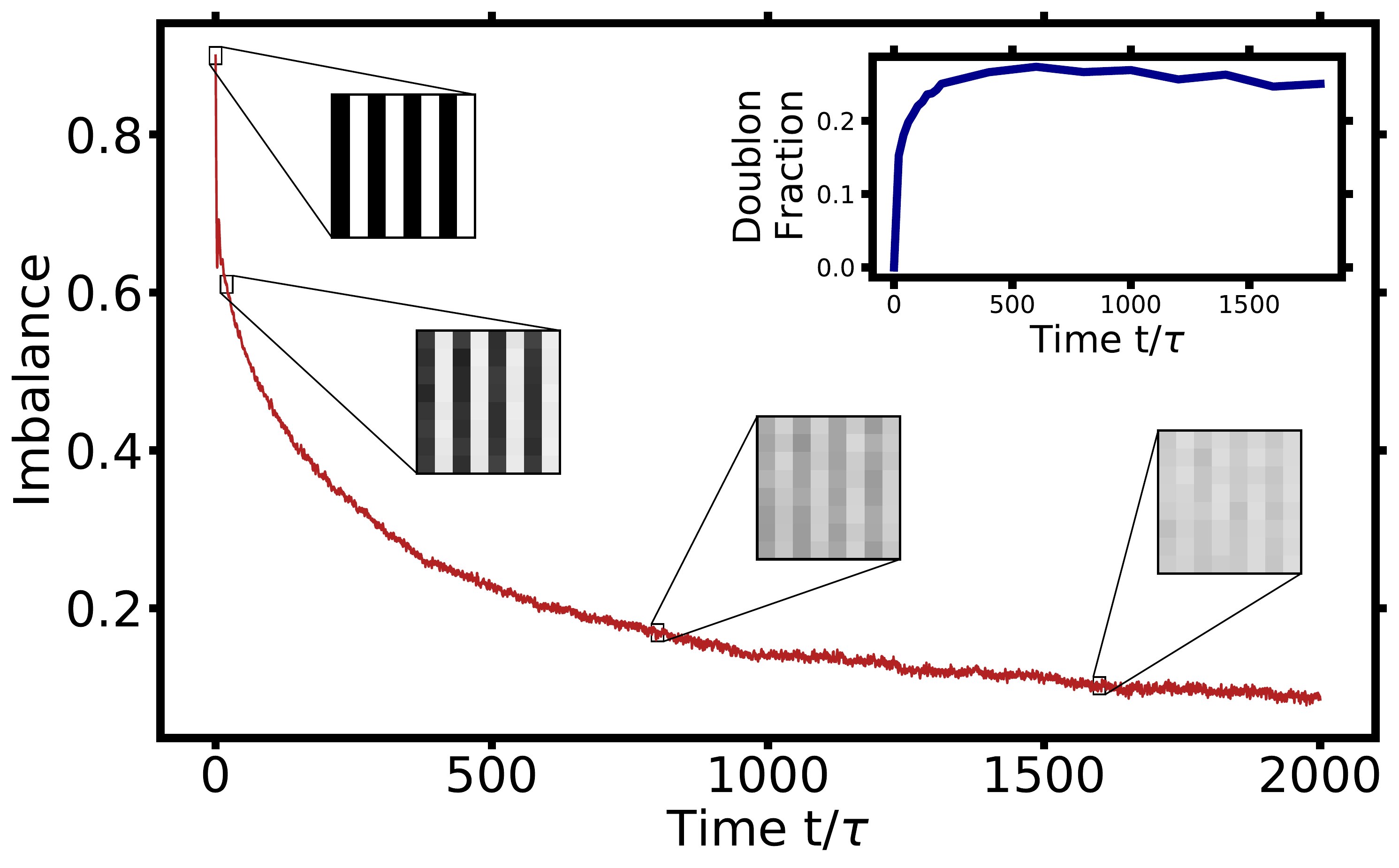}
    \caption{(Color Online)
    Time evolution of the imbalance. The system contains 31 dirty particles (which feel a disordered potential)  and a single clean particle (which does not experience the disorder) on an 8x8 lattice. The imbalance, $I=(N_{\rm even}-N_{\rm odd})/(N_{\rm even}+N_{\rm odd})$, is a proxy for how far the system is from equilibrium. $N_{\rm even}$ and $N_{\rm odd}$ are the number of particles on alternate columns of the lattice. Each point represents the average over 216 disorder realizations. Time is measured in units of $\tau=\hbar/J$, where $J$ is the single-particle tunneling rate. Call-outs show the spatial distribution of dirty particles at: $t=$ 0$\tau$, 100$\tau$, 800$\tau$, 1600$\tau$. Darker colors represent higher densities. Inset: Dynamics of the doublon fraction, defined as the ratio of particles on doubly-occupied sites to the total number of particles. %The standard error in the mean is too small to be visible in the inset plot.
    }
    \label{Fig:F=1}
\end{figure}

In Fig.~\ref{Fig:F=1}, we present numerical results for an 8x8 lattice, with periodic boundary conditions. This consists of 32 particles in total. For this figure, we average over 216 disorder realizations. To the extent that a sufficiently large system is self-averaging, this ensemble averaged result is indicative of the physics in the thermodynamic limit. Our system size is somewhat smaller than the 124 particles in the experiment \cite{Main}, where results are averaged over approximately 50 realizations. Despite these parameter differences, our results are very similar to the experimental observations.  This is consistent with the fact that our results do not change significantly when we use different system sizes.

\begin{figure*}[tbph]
    \centering
    \includegraphics[width=\textwidth]{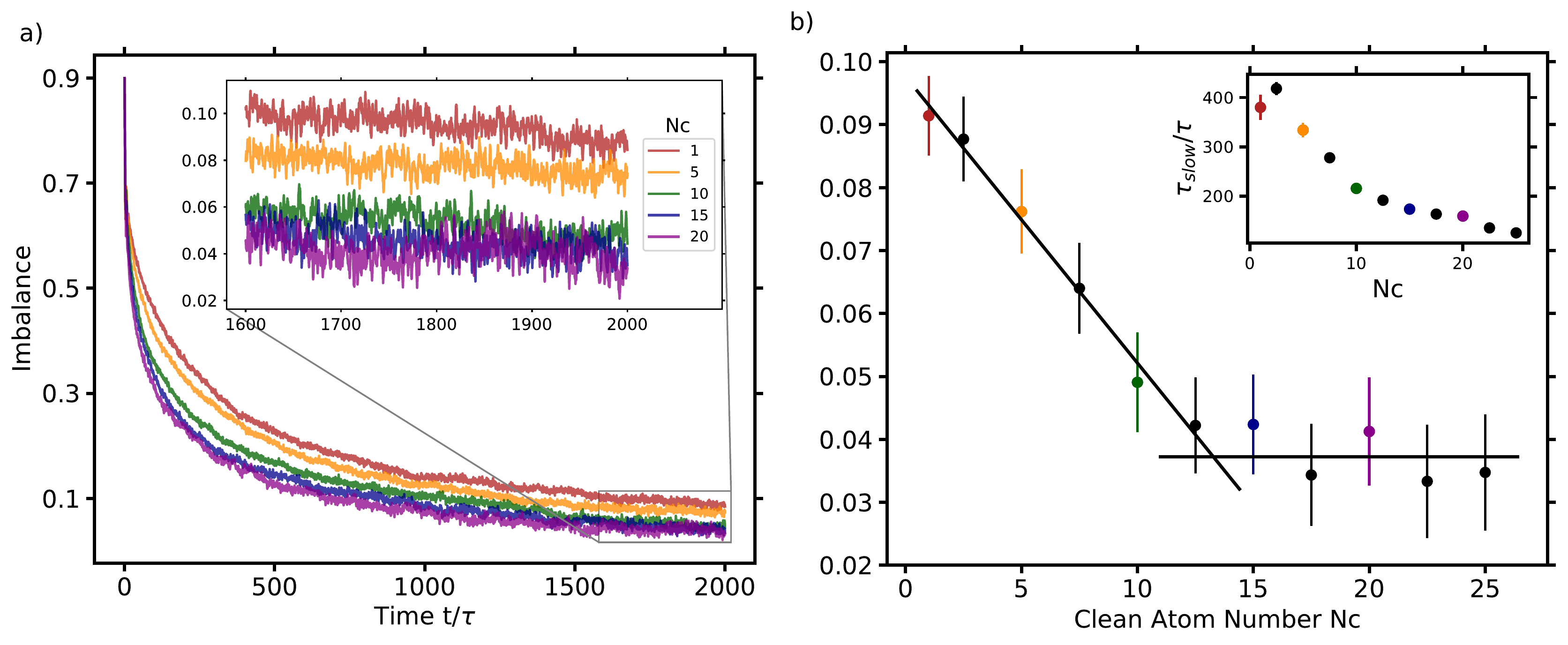}
    \caption{(Color Online) Effect of the bath size on imbalance dynamics. The system has a total of $N_c+N_d=32$ particles, and 64 sites, where $N_c$ is the number of ``clean"  particles, which form the bath and are not influenced by the disorder, while $N_d$ is the number of ``dirty" particles which experience the disorder potential. a) Imbalance versus time: from top to bottom (red, orange, green, blue, purple): $N_c=1,5,10,15,20$, $N_d=31,27,22,17,12$. Each point corresponds to the average of 216 disorder realizations. Inset: Zoomed in plot of behavior at long times. b) Final imbalance (average imbalance between times $t=1800\tau$ and $2000\tau$) versus number of clean particles. Error bars represent the standard error of the mean. The ``threshold" behavior is highlighted by the guides to the eye. Inset: Time, $\tau_{slow}$, at which the imbalance falls to $0.7/e=0.26$.}
    \label{Fig:Final Imbalance}
\end{figure*}

The imbalance shows features on three different timescales. Initially, a sharp decrease is observed on a timescale $\tau_{\rm fast}$ with $\tau_{\rm fast}\sim \tau=\hbar/J$. This rapid decay is due to the hopping of particles into neighboring empty sites. On the scale of the figure, this initial evolution appears as a nearly vertical line. Subsequently, the imbalance shows a much slower decay, on a timescale $\tau_{\rm slow}$ with $\tau_{\rm slow}\sim 400 \tau$. This slow decay involves more significant spatial rearrangement of the particles. At longer times ($t>1500\tau$) the imbalance continues to fall, but much slower still: to first approximation it is constant. A visual representation of the imbalance is seen in the call-outs, which show the average density of dirty particles in the system, with darker shades representing higher particle numbers. The contrast of the fringes decreases, although never vanishes, on the time-scale of the simulations. The other relevant feature in Fig.~\ref{Fig:F=1}, is the high frequency noise, which is on the scale of $\tau$.  It comes from the rapid motion of atoms between neighboring sites. 

The experimentalists also study the time dynamics of the fraction of doubly occupied sites. The inset of Fig.~\ref{Fig:F=1} shows the fraction of particles in doubly occupied sites from our numerics. This fraction grows over a timescale of  $\sim 100\tau$, then stays roughly constant at around 24$\pm$4$\%$. This behavior is similar to what is seen in the experiment. There is no obvious connection between the doublon dynamics and the imbalance.

Although not shown on this graph, the imbalance of clean particles decays quickly to zero on a timescale of a few $\tau$. This is consistent with the interpretation of the clean particles as a bath, and is consistent with the observations of the experiment.

\subsection{Dependence of Final Imbalance on Bath size}

Figure~\ref{Fig:Final Imbalance} a) shows the time dependence of the imbalance for various fractions of clean particles. The curves have the same general shape as Fig.~\ref{Fig:F=1}.  The most striking difference is in the long-time behavior, illustrated in the inset. To quantify the long-time imbalance we average between $t=1800\tau$ and 2000$\tau$, which is the longest timescale accessible by the experiment.  As shown in Fig. \ref{Fig:Final Imbalance} b), there is a ``threshold" behavior.  For small $N_c$, the long-time imbalance decreases linearly with $N_c$. For large $N_c$ the imbalance is independent of $N_c$.  The threshold separating these behaviors is at approximately $N_c^*=13$, corresponding to a clean fraction of 40\%.  The experiment observes similar results. 

One significant difference between our numerics and the experiment is that our long-time imbalance is non-zero at large $N_c$; in the experiment \cite{Main}, the long-time imbalance reaches zero. There the threshold is taken to indicate a phase transition from a many body localized state to an ergodic ``thermalized" state.

Our observed long-time imbalance is highly sensitive to the disorder realization.  The error bars in Fig.~\ref{Fig:Final Imbalance} shows the standard error of the mean -- the standard deviation is nearly 15 times larger.  

As we described in Sec.~\ref{sec:id}, the imballance dynamics occurs on several timescales.  The timescale of fast decay to $I \sim 0.7$ is independent of $N_c$, and is of order $\hbar/J$.  We quantify the slow timescale, $\tau_{\rm slow}$, by finding the time at which $I=0.7/e=0.26$.  The inset of Fig.~\ref{Fig:Final Imbalance} b) shows the dependence of $\tau_{\rm slow}$ on $N_c$.  It does not have the distinct threshold behavior seen in the long-time imbalance, but there may be a break in the slope near $N_c^*$.  Larger baths result in more rapid decay.

\begin{figure*}[tbph]
    \centering
    \includegraphics[width=\textwidth]{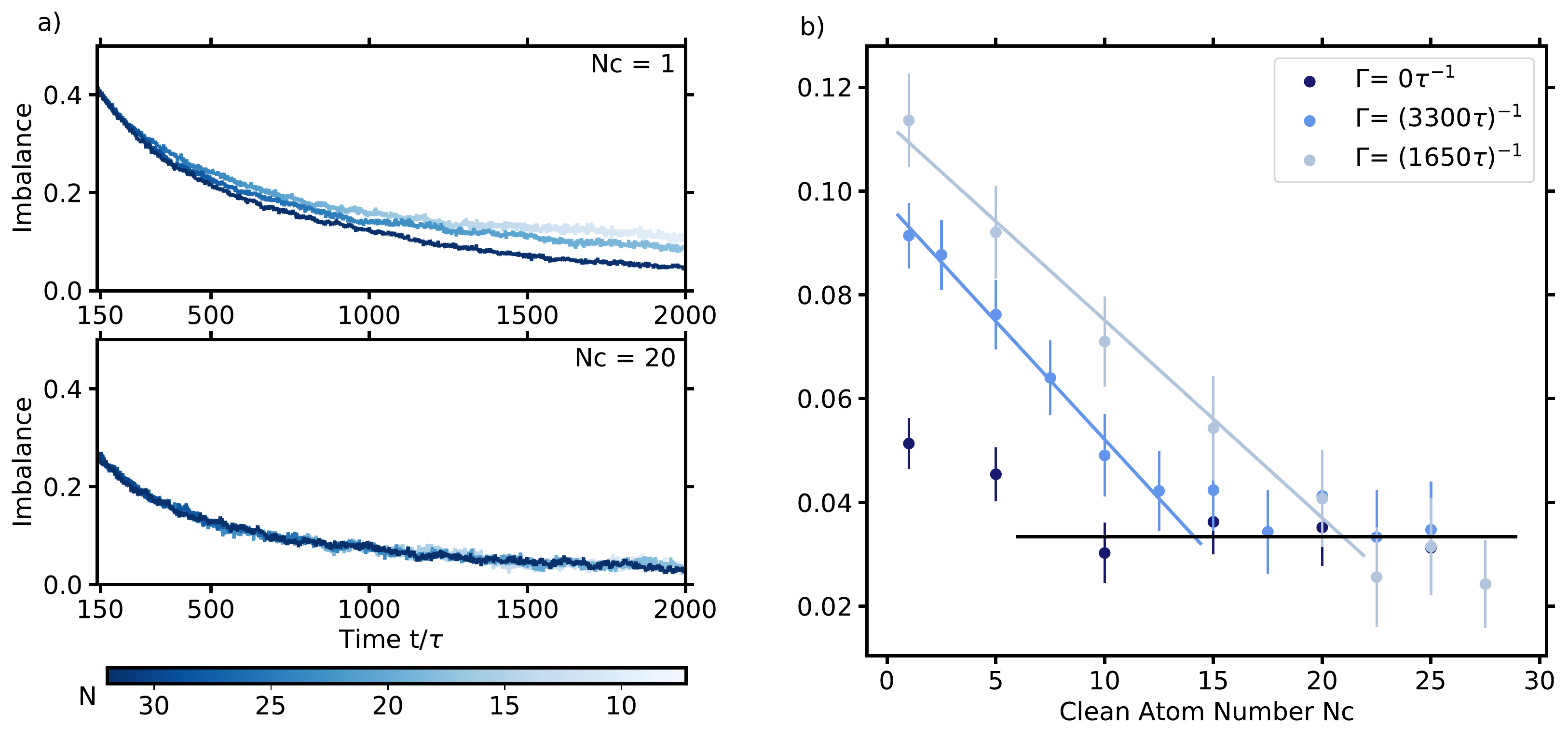}
    \caption{(Color Online)  Effect of particle loss on imbalance. a) Time evolution: Top -- small bath, $N_c = 1, N_d = 31$; Bottom -- large bath, $N_c = 20, N_d = 12$. Three loss rates are shown, dark to light: $\Gamma=0$,$\Gamma_0$, $2\Gamma_0$, where  
    $\Gamma_0=(3300\tau)^{-1}$ is the experimentally observed rate. Shading denotes the total number of particles ($N=N_c+N_d$) still remaining in the system. b) Final imbalance (average imbalance between times $1800\tau$ and $2000\tau$) versus number of clean particles.  The threshold moves to smaller $N_c$ as the loss rate is reduced. For the scenario with no loss, the threshold is no longer prominent.
    }
    \label{Fig:Loss}
\end{figure*}

\subsection{Dependence of Final Imbalance on Loss rate}

Surprisingly, we find that particle loss plays a crucial role in the behavior of the system. In the plots of Fig.~\ref{Fig:Final Imbalance}, we have set the loss rate to that observed in the experiment, $\Gamma = (3300 \tau)^{-1}.$ In Fig.~\ref{Fig:Loss}, we compare those results to scenarios with either no particle loss, or twice as large of a loss rate. For systems with a small bath,  as illustrated with $N_c = 1, N_d = 31$, the long-time imbalance is visibly different for the three different loss scenarios. However when the size of the bath is relatively large, as illustrated with $N_c = 20, N_d = 12$, the curves lie on top of each other for all times.  In particular, their final imbalances differ by less than the standard error in the mean. Fig.~\ref{Fig:Loss} b) compares the dependence of the final imbalance on the number of clean particles for the three different cases. In the absence of loss the final imbalance is nearly independent of $N_c$. If there is a threshold behavior, it is subtle. As previously described, when the loss rate is tuned to the experimental value, we find a characteristic cross-over between linear and constant regimes. Doubling $\Gamma$ from $(3300\tau)^{-1}$ to $(1650\tau)^{-1}$ pushes the threshold from $N_c = 13$ to $N_c = 21$. The large $N_c$ results are independent of the strength of particle loss. 

Due to these theoretical observations, it is imperative to repeat the experiment with different loss rates. Reducing loss is hard, but it is readily  increased by adjusting the background pressure in the vacuum chamber.  One can also tune the optical lattice frequency closer to resonance, leading to enhanced off-resonant light scattering.  The important role of atom loss may make it harder to interpret the experimental results in terms of many-body localization.

\section{Summary and Outlook}

We modelled the dynamics of a two-component Bose gas in the presence of a random on-site disorder which affects only one of the components. We find that an initially prepared period two density wave decays with two timescales:  On the time-scale of the hopping, there is a sudden drop in the imbalance. Subsequently, there is a much slower decay, which is due to an interplay between interactions, disorder, and atom loss. As seen in the experiment \cite{Main}, we find that the long-time imbalance depends on the number of clean atoms, $N_c$, which act as a bath. The size of the bath no longer matters once it exceeds a certain threshold.  
 
We find that particle loss is required to see this behavior. To model this loss we used a stochastic wavefunction approach, together with our variational ansatz. The impact of the bath grows as particle loss is increased. In future experiments extra loss can be engineered to explore this phenomenon.

\section{Acknowledgements}
This work was supported by the NSF Grant PHY-1806357.  SP further acknowledges support from the Herchel Smith Scholarship funded by Emmanuel College.

\appendix

\section{Time Stepping}\label{lagrange}

\subsection{Overview}\label{lagrangeoverview}
We use a split-step method to evolve the coefficients of the variational wavefunction, separately considering the Hermitian and non-Hermitian terms.  This separation allows us to more carefully control any numerical errors which could erroneously lead to atom loss.  

For the Hermitian component, we use a Runge-Kutta method (RK4), to evolve Eq.~(\ref{master}) for one timestep. Small numerical errors, of order $\mathcal{O}(\delta t ^5)$, will introduce some artificial loss. Because the loss in the experiment is extremely small, we need to ensure that these numerical errors do not accumulate and lead to erroneous results.  We correct for this numerical loss by using the renormalization procedure in Appendix \ref{nc}. We then explicitly introduce the physical loss, evolving the coefficients via the Euler method.  The Euler method is sufficient, as errors are on the order of $\mathcal{O}(\Gamma \delta t ^ 2)$, and $\Gamma$ is very small.   Finally, as described in Appendix~\ref{atom_loss} we include a stochastic step.

\subsection{Enforcing Number Conservation}\label{nc}
Before each step we calculate the total number of clean and dirty particles, $N_c^*$, $N_d^*$. We then use a single step of the RK4 method on Eq.~(\ref{master}) to account for the Hermitian evolution. The coefficients in our time-evolved wavefunction are $f^i_{n_c,n_d}$.  We wish to find new coefficients $g^i_{n_c,n_d}$, which are as close as possible to $f^i_{n_c,n_d}$, while satisfying the constraints
\begin{eqnarray}
\sum_{i,n_c,n_d}n_c |g^i_{n_c,n_d}|^2&=&N_c^*\nonumber\\
\sum_{i,n_c,n_d}n_d |g^i_{n_c,n_d}|^2&=&N_d^*\nonumber\\
\sum_{n_c,n_d}|g^i_{n_c,n_d}|^2&=&1.
\end{eqnarray}
We perform this optimization by 
minimizing the Lagrangian
\begin{widetext}
\begin{multline}
    L = \sum_{i,n_c,n_d} |f^i_{n_c,n_d} - g^i_{n_c,n_d}|^2 + \mu_{c}(\sum_{i,n_c,n_d} n_{c}|g^i_{n_c,n_d}|^2 - N_{c}^{*}) + \mu_{d}(\sum_{i,n_c,n_d} n_{d}|g^i_{n_c,n_d}|^2 - N_{d}^{*}) +  \sum_{i}\alpha_{i}(|g^i_{n_c,n_d}|^2 - 1),
\end{multline}
\end{widetext}
where $\mu_c$, $\mu_d$ and $\alpha_i$ are lagrange multipliers.  Due to the small magnitude of constraint violations, the Lagrange multipliers are small. To first order in them, the minimization with respect to $g_{n_c,n_d}^*$ yields
\begin{equation}\label{eqg}
g^i_{n_c,n_d} = f^i_{n_c,n_d}(1 + \mu_{c}n_{c} + \mu_{d}n_{d} + \alpha_{i}).
\end{equation}
Defining $N_c = \sum_{i,n_c,n_d} n_{c}|f_{i,n_c,n_d}|^2$, and $N_d = \sum_{i,n_c,n_d} n_{d}|f_{i,n_c,n_d} |^2$, we compute the variances and covariances: $\Delta_{cc} = \sum_{i} \langle n_{c,i}^2 \rangle - \langle n_{c,i} \rangle ^2, \Delta_{dd} = \sum_{i} \langle n_{d,i}^2 \rangle - \langle n_{d,i} \rangle ^2, \Delta_{cd} = \sum_{i} \langle n_{c,i}n_{d,i} \rangle - \langle n_{c,i} \rangle \langle n_{d,i} \rangle$, in terms of which the Lagrange multipliers are:
\begin{eqnarray}
    \begin{bmatrix}
    \mu_c \\ \mu_d
    \end{bmatrix}
     &=&
    \begin{bmatrix}
    \Delta_{cc} & \Delta_{cd} \\
    \Delta_{cd} & \Delta_{dd}
    \end{bmatrix}^{-1}
    \begin{bmatrix}
    (N_{c}^* - N_{c})/2 \\
    (N_{d}^* - N_{d})/2
    \end{bmatrix}\\
    \alpha_i &=& -\mu_{c}n_{c,i} - \mu_{d}n_{d,i}.
\end{eqnarray}
We use these results to calculate $g^i_{n_c,n_d}$ in Eq.~(\ref{eqg}).

\section{Atom Loss}\label{atom_loss}
We use the stochastic wavefunction approach from \cite{Stochastic-Loss} to model atom loss. We assume that every particle has an equal probability of being lost, as would be appropriate for background-gas collisions or inelastic light scattering. The experiment observes similar loss rates for both components, characterized by a single value $\Gamma=1/3300\tau$. 

The stochastic wavefunction approach involves interleaving coherent time evolution with random ``quantum jumps", and averaging over the resulting ensemble. We calculate these dynamics within the Gutzwiller Ansatz. As described in Appendix~\ref{lagrangeoverview}, we use a split-step approach for the coherent dynamics, first evolving with the Hermitian terms in Eq.~(\ref{master}). We then evolve with the non-Hermitian Hamiltonian in Eq.~(\ref{loss}).  Within the Euler approximation, this  results in a stepping rule
\begin{equation}\label{loss_tdse}
 f_{n_{c},n_{d}}^{i} \to
 \left(1-\frac{\delta t \Gamma}{2}(n_{c} + n_{d})\right)f_{n_{c},n_{d}}^{i}.
\end{equation}
The probability that we have lost a clean particle from site $i$ during that time step is
\begin{equation}
\delta p_{c,i} = -\delta t \Gamma  \sum_{n_c,n_d} n_c |f_{n_c,n_d}^i|^2,
\end{equation}
and a similar expression gives the probability of loosing a dirty particle, $\delta p_{d,i}$. We use a random number generator to check to see if any of these events occurs.  If a  particle is lost at site $i$ we take $|\psi\rangle\to \hat a_{ci}|\psi\rangle$ or $|\psi\rangle\to \hat a_{di}|\psi\rangle$.  Whether or not a loss event occurs, we then normalize the wavefunction. 

%\bibliography{bibliography.bib}

\end{document}